\documentclass[eqsecnum,floats,aps,pra,floatfix,titlepage,tightenlines]{revtex4} 

\usepackage{graphicx}
\usepackage{graphics}
\usepackage{bm}
\usepackage{amssymb}
\usepackage{amsmath}
\usepackage{mathrsfs}

\begin{document}

\title{Frequency Spectra of Isolated Laser Pulse Envelopes}

\author{L. H. Ford}
 \email{ford@cosmos.phy.tufts.edu}
 \affiliation{Institute of Cosmology, Department of Physics and Astronomy, Tufts University, Medford, Massachusetts 02155, USA}
  
  \author{Brian Fu}
 \email{Brian.Fu@tufts.edu}
 \affiliation{Institute of Cosmology, Department of Physics and Astronomy, Tufts University, Medford, Massachusetts 02155, USA}
  
\begin{abstract}
This paper will deal with isolated laser pulses, those which last for a finite time interval and whose envelope function is strictly zero outside
of this interval. We numerically calculate the Fourier transform of this function and study its asymptotic behavior at high frequencies. This work
is motivated by recent results on the probability distributions of quadratic operators in second quantized systems. An example is the density of
a material which is subject to zero point fluctuations in the phonon vacuum state. These distributions can decrease very slowly, leading to a relatively
high probability  for large fluctuations. If the operator is measured by a laser pulse, the rate of decrease of the distribution mirrors the rate of decrease  
of the pulse envelope Fourier transform. We describe a model for the creation of isolated pulses in which this Fourier transform falls as an exponential 
of a fractional power of frequency and find examples where this fraction is in the range 0.1 to 0.2. The probability distribution for large fluctuation has the 
same functional form and implies a significant probability for fluctuations very large compared to the variance.
\end{abstract}

\maketitle

\baselineskip=14pt

\section{Introduction}
\label{sec:intro}

The primary topic of this paper will be  the envelope function of an isolated laser pulse of finite duration. Such a pulse may be viewed as a wave packet in
time of the form
\begin{equation}
 G(t) = f(t) \, s(t)\,.
 \end{equation}
Here $s(t)$ is an approximately sinusoidal signal function, and $f(t)$ is a much more slowly varying envelope function with the property that it vanishes 
outside of a finite time interval. This describes the finite duration of the pulse. Our main concern will be with the Fourier transform, $\hat{f}(\omega)$, of
the envelope function
\begin{equation}
 \hat{f}(\omega) = \int_{-\infty}^\infty dt \, f(t) \, {\rm e}^{i \omega t} \,,
 \end{equation}
and especially with its asymptotic form for large $\omega$.

A function, such as $f(t)$, which vanishes outside of a finite interval is known in mathematics as having compact support. An infinitely differentiable function
with compact support necessarily has a Fourier transform which falls more slowly than any exponential but faster than any power at high frequencies, 
as discussed in Appendix~\ref{sec:App A}. 
If $\tau$ is the approximate temporal duration of $f(t)$, then we may take
high frequency to mean $ \tau \, \omega \gg 1$. An example of a function which falls slower than an exponential but faster than any power is one with
the asymptotic form
\begin{equation}
 |\hat{f}(\omega)| \sim  {\rm e}^{- a\, (\tau \, \omega)^\alpha} \,, \quad  \tau \, \omega \gg 1 \,,
 \label{eq:FT-asy}
 \end{equation}
where $a >0$ and $0 < \alpha < 1$.

The physical significance of this asymptotic form lies in its link to the asymptotic probability distribution for large fluctuations of quadratic operators in
second quantized theories~\cite{FF15,SFF18,FF2020,WSF21}. First consider  a relativistic quantum field, such as electromagnetism.  A field operator at a single spacetime point is not
an observable in the sense that its moments diverge, but an average over a finite region of space and time has finite moments and may be meaningful.
The effect of the spacetime averaging is to suppress the contributions of very high frequency modes. In the case of a quadratic operator, such as the
energy density, the averaging must include a time average, but a spatial average is optional.  We may regard the space and time averaging as being
associated with a measurement of an observable, such as the averaged energy density in a finite region.

Let $v(t)$ be a quadratic operator, which could be either at one space point or a spatial average. Now define its time average, $u$, with respect to some
compactly supported function $f(t)$:
 \begin{equation}
 u =\frac{\int_{-\infty}^\infty dt \, f(t) \, v(t)}{\int_{-\infty}^\infty dt \, f(t)}\,.
 \end{equation}
Now all of the moments of $u$ in any quantum state are finite: $\langle u^n \rangle < \infty$ for all positive integers. The eigenvalues of the averaged
operator $u$ may be viewed as possible outcomes of a measurement of the averaged energy density, for example, in a finite region of spacetime.

A key result, discussed in Refs.~\cite{FF15,SFF18,FF2020,WSF21}, is that the probability distribution for a given outcome $u$ has the same asymptotic form as does the Fourier 
transform of $f(t)$, 
 \begin{equation}
 P(u) \sim c_0\, {\rm e}^{- b\, u^\alpha } \, \quad  u \gg 1 \,.
 \label{eq:Pu}
 \end{equation}
Here we assume that $u$ is dimensionless, scaled to have a variance of one, and $c_0$ and $b$  are constants.  We may also give the
asymptotic  complementary cumulative distribution, which is the probability to find an outcome at least as large as $u$ :
\begin{equation}
 P_>(u)  = \int_u^\infty dy \, P(y)   \sim \frac{c_0}{\alpha\, b}\, u^{1 - \alpha}\,   {\rm e}^{- b\, u^\alpha } \,, \quad  u \gg 1 \,.
 \label{eq:P-greater}
 \end{equation}

This result implies a relatively 
high probability to observe large fluctuations, those much larger than described by the variance. Furthermore, this probability increases significantly
for smaller values of $\alpha$. It was shown in Ref.~\cite{WSF23}  that a given region of the asymptotic distribution Eq.~\eqref{eq:Pu} is determined
by modes whose angular frequencies are of order of a dominant frequency
 \begin{equation}
 \omega_d \approx \frac{u}{ \tau}\,. 
 \label{eq:omega-d}
 \end{equation}
Thus the exponential in the  asymptotic functional forms of $P(u)$ and $P_>(u)$  mirrors that in the asymptotic Fourier transform of the averaging function, $\hat{f}(\omega)$.
However, the constants $c_0$ and $b$ can depend upon other details of the averaging function, as well as the choice of the specific operator $v(t)$ being measured.
 
A main purpose of the present paper is to explore how small $\alpha$ could be when the averaging is implemented by
an isolated laser pulse. This will provide insights into the probability of large quantum fluctuations of quadratic operators measured by this pulse.
The outline of this paper is as follows: Section~\ref{sec:rate} will present  a simple model for the generation of an isolated output laser pulse using
rate equations. Some sample numerical solutions for the envelope function will be discussed. In Sect.~\ref{sec:Fourier}, we treat the Fourier transforms
of these solutions and find that in limited ranges at high frequency, they can have the form of Eq.~\eqref{eq:FT-asy}, and find the associated values of the
parameter $\alpha$.  Some applications of the results to specific physical systems will be given in Sect.~\ref{sec:apps}.    
Our results will be summarized and discussed in Sec.~\ref{sec:final}.

\section{The Rate Equations}
\label{sec:rate}

Our model describes a single shot laser which is pumped by an incoming isolated photon pulse. Here $x(t)$ is the number of photons
in a particular cavity mode at time $t$, and $a(t)$ is the number of atoms in the excited state at that time. Both are assumed to vanish
before the arrival of the pump pulse: $x(0) = a(0) = 0$.  The equation for $x(t)$ is

 \begin{equation}
 \frac{dx}{dt} = C\, [x(t) +1]\, a(t) -\eta \, C\, [N_0 - a(t)]\, x(t) -r\, x(t) \,.
 \label{eq:x-eq}
 \end{equation}
 The first term on the right hand side is the rate at which photons are emitted into the selected mode by spontaneous and stimulated emission.
 Thus the constant $C = 1/\tau_{\rm life}$, where $\tau_{\rm life}$ is the radiative lifetime of the excited state due to spontaneous  emission.
 The second term, proportional to the parameter $\eta$, is the rate at which the photons are re-absorbed by an atom in the cavity.  Here we assume 
 $0 \leq \eta \leq 1$ and focus on two limiting cases, $\eta = 0$, no  re-absorption, and $\eta = 1$, strong  re-absorption.   The final term in Eq.~\eqref{eq:x-eq}
 is the rate at which photons leave the cavity through a partially reflecting mirror. Thus $r\, x(t)$ is the photon number flux in the outgoing pulse. 
 The equation for $a(t)$ is
  \begin{equation}
 \frac{da}{dt} =  N_0\, f_\nu(t) - C\, [x(t) +1]\, a(t) +\eta \, C\, [N_0 - a(t)]\, x(t)  \,.
  \label{eq:a-eq}
 \end{equation}
The second and third terms on the right hand side are minus the corresponding terms in Eq.~\eqref{eq:x-eq}, which describe the effects of photon 
emission and re-absorption. The first term is the effect of the pump pulse.

\subsection{The Pump Pulse}
\label{sec:pump}

We assume that the photon number flux in the pump (input) pulse is proportional to a compactly supported function of time, $f_\nu(t)$ with unit time integral:
  \begin{equation}
 \int dt \,  f_\nu(t) = 1\,.
 \end{equation}
 Thus the constant $N_0$ in Eq.~\eqref{eq:a-eq} is the number of atoms which would be excited by the pump pulse ignoring emission and re-absorption  effects,
 that is, in the $C = 0$ limit. 

 We will assume a particular class of compactly supported functions, which were discussed in Ref.~\cite{FF15}, and defined by
  \begin{equation}
 f_\nu(t) = k_\nu \; \exp\left[ - \frac{2}{[t (1-t)]^\nu} \right]
 \end{equation}
where $\nu > 0$ and $ k_\nu$ is a normalization constant. First note that  $f_\nu(t)$ vanishes faster than any power as $t \rightarrow 0^+$ and
 $t \rightarrow 1^-$. Hence, we may take $f_\nu(t) = 0$ outside the finite interval $0 < t < 1$. Note that we are adopting units in which the duration of the pump pulse 
 has been set to unity, $\tau_{\rm in} = 1$.
 
 We may choose the duration of the pump pulse, $\tau_{\rm in}$, to have any value in conventional units. By adopting units in which
$\tau_{\rm in} = 1$, we are making all of the quantities in Eqs.~\eqref{eq:x-eq} and \eqref{eq:a-eq} dimensionless; all times are multiples of
$\tau_{\rm in}$ and all rates are multiples of $1/\tau_{\rm in}$.

\begin{figure}[htbp]
\includegraphics[scale=0.2]{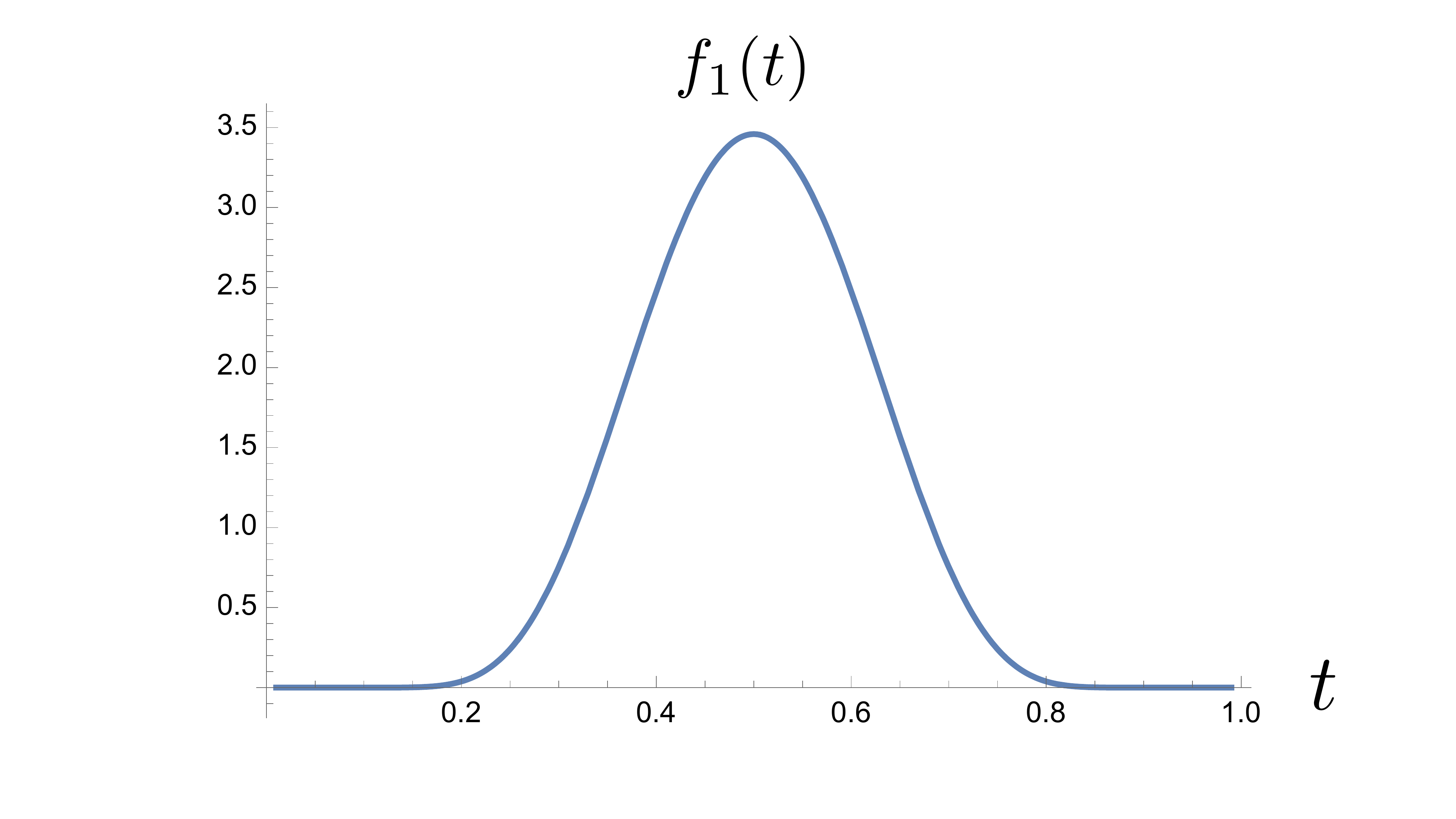}
\caption{ The function $f_1(t)$  is illustrated. Note that although $f_1(t) \not= 0$ for $0 < t < 1$, its actual value is small outside of
$0.2 \alt t \alt 0.8$.}
\label{fig:f1}
\end{figure}

The Fourier transform,  $ \hat{f}_\nu(\omega)$,  decays as the exponential of a fractional power for large $\omega$, as in Eq.~\eqref{eq:FT-asy},
where 
\begin{equation}
 \alpha = \alpha_{\rm in} = \frac{\nu}{1+\nu}\,.
 \end{equation}
We are particularly interested in the special case $\nu = 1$, or $\alpha_{\rm in} = \frac{1}{2}$. This case describes a simple electrical circuit
in which  the  current rises as $f_1(t)$ just after  a switch is closed. [See Sect.~IIC of Ref.~\cite{FF15}.]  In this case, $k_1 \approx 1.031 \times 10^4$, and
$f_1(t)$ is plotted in Fig.~\ref{fig:f1}

\subsection{Examples of the output pulse}
\label{sec:output1}

\begin{figure}[htbp]
\includegraphics[scale=0.2]{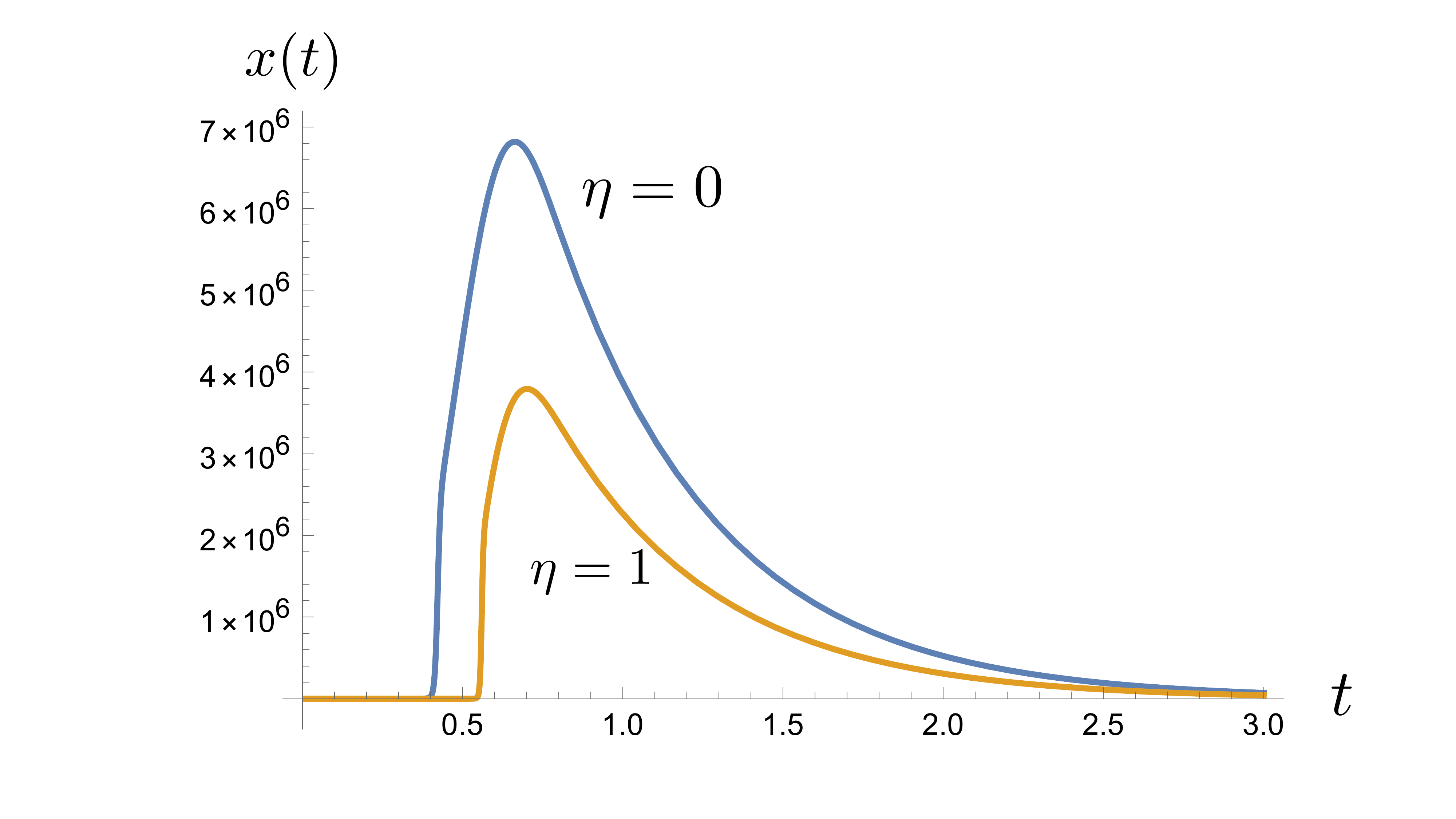}
\caption{The photon number for the case $N_0 = 10^7$, $C=10^{-4}$, $r = 2$ is plotted with no re-absorption, $\eta = 0$, and for strong re-absorption,
$\eta = 1$}
\label{fig:photon-no-1}
\end{figure}

Here we present two examples of numerical solutions of  Eqs.~\eqref{eq:x-eq} and \eqref{eq:a-eq} for the case $N_0 = 10^7$, $C=10^{-4}$, and 
$r=2$. We consider two choices for the re-absorption parameter: $\eta = 0$, no re-absorption, and  $\eta = 1$, strong re-absorption. The results 
for the photon number, and hence the output pulse, are plotted in Fig.~\ref{fig:photon-no-1}.
 We see that the effect of  re-absorption is to slightly delay the onset of the output pulse and to reduce its magnitude. Otherwise the qualitative 
 features are not noticeably altered. In both cases, the output pulse is asymmetric, with a sharp rise and slower decay in time. Its temporal
 duration, $\tau_{\rm out}$, is somewhat longer than that of the pump pulse, but of the same order of magnitude. Roughly, $\tau_{\rm out} \approx 2 \, \tau_{\rm in} $.
 \begin{figure}[htbp]
\includegraphics[scale=0.2]{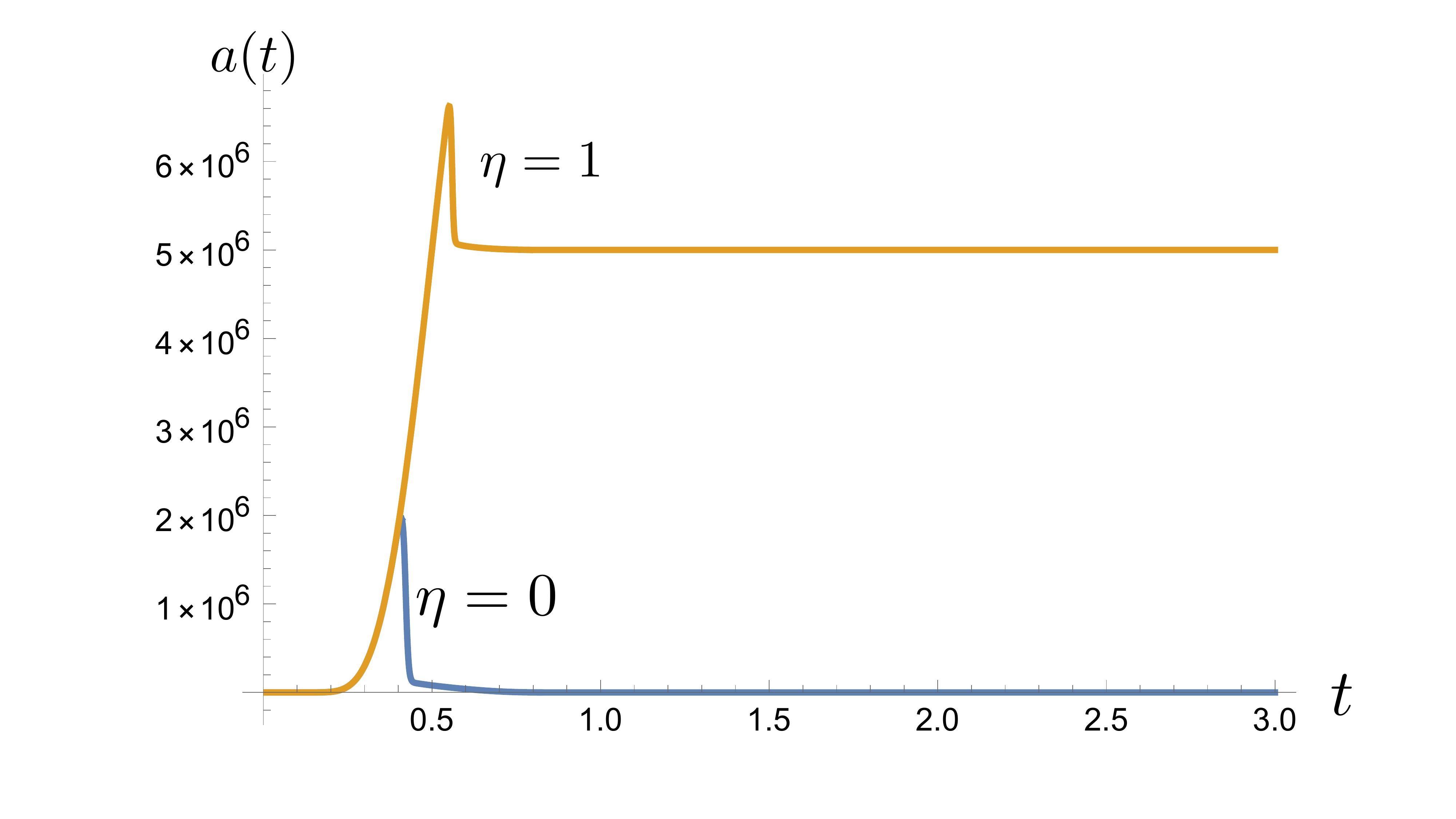}
\caption{The  number of excited atoms is plotted for the same choices of parameters as  in Fig.~\ref{fig:photon-no-1}. } 
 \label{fig:atom-no-1}
\end{figure}  
The number of atoms in the excited state as a function of time is plotted in Fg.~\ref{fig:atom-no-1}. Here we see a marked difference between the cases of no 
re-absorption ($\eta = 0$), and the case $\eta = 1$. In the former case, $a(t)$ spikes and then quickly approaches zero. In the latter case, $a(t)$  decreases
to a finite nonzero value which remains constant on the scale of this plot. We can interpret this behavior as follows: Re-absorption not only allows $a(t)$ to reach 
a greater maximum value, but also causes it to reach a nonzero value after nearly all of the photons have left the cavity. At this point, stimulated emission ceases,
although $a(t)$ continues to decrease slowly through spontaneous emission. 

\begin{figure}[htbp]
\includegraphics[scale=0.2]{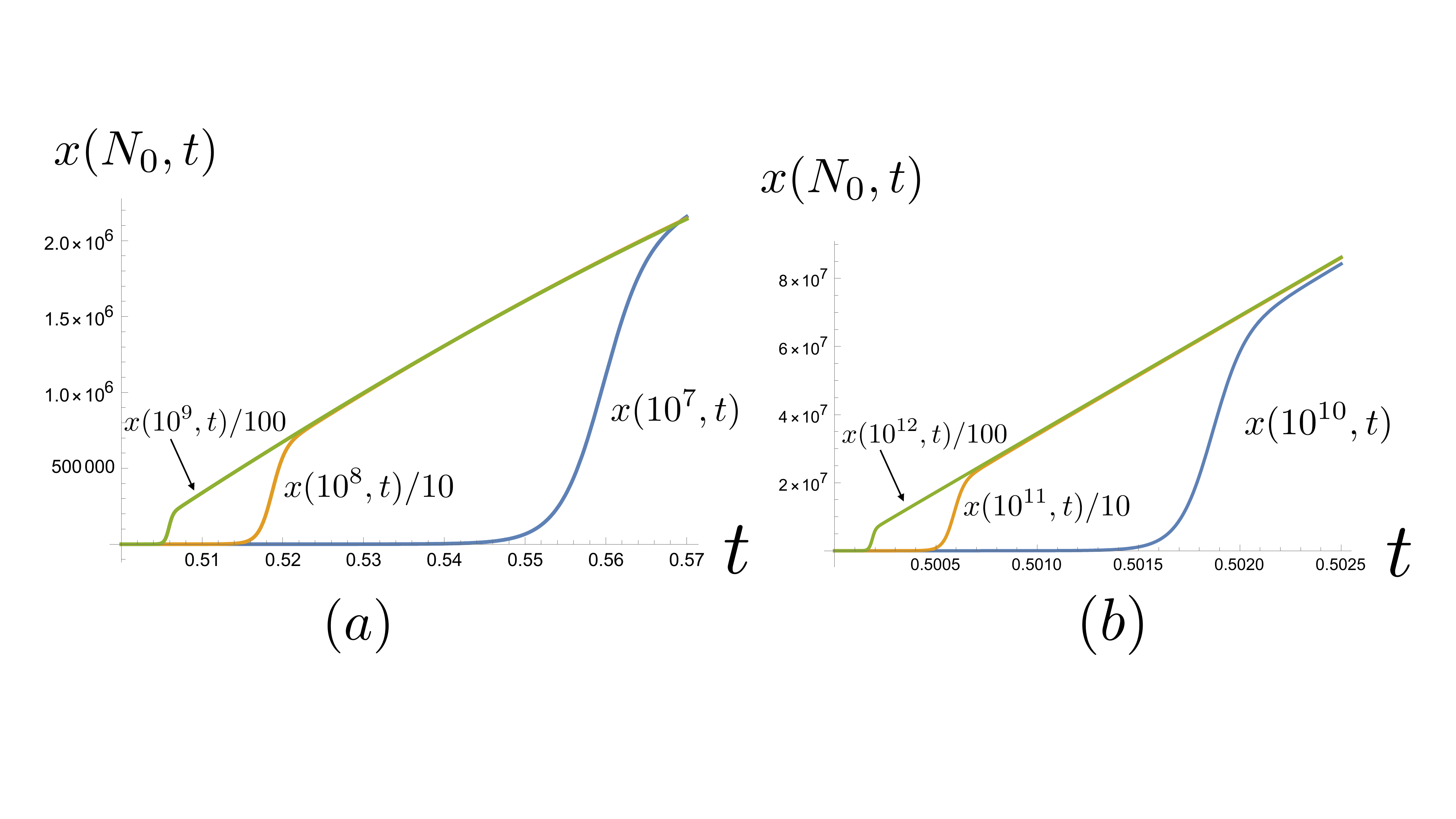}
\caption{The behavior of the photon number, $x(N_0, t)$, as a function of time is illustrated near the switch-on time for different values of $N_0$. Part (a) shows the cases
$N_0 = 10^7, 10^8,$ and  $10^9$. Because the photon number scales with $N_0$, the cases $N_0 = 10^8$ and $N_0 = 10^9$ have been rescaled by factors of $\frac{1}{10}$ and
$\frac{1}{100}$, respectively, so that all three cases could be shown on the same plot. Part (b) gives a similarly rescaled plot for the cases $N_0 = 10^{10}, 10^{11},$ and  $10^{12}$.
Note that as $N_0$ increases, the initial switch-on time moves closer to $t= 0.5$, the midpoint of the pump pulse. Otherwise, the rescaled graphs all have the same form. In all the cases 
illustrated here, we have set $\eta = 1$.}
\label{fig:switch-on}
\end{figure}

We see in Fig.~\ref{fig:photon-no-1} that the photon number in the cavity, $x(t)$, rises rapidly from zero to a maximum value somewhat less than $N_0$. The time at
which this sharp rise begins depends both upon the re-absorption parameter, $\eta$, and upon the photon numner parameter, $N_0$. In the case illustrated, $N_0 = 10^7$, 
this time is $t \approx 0.4$ for  $\eta = 0$ and $t \approx 0.55$ for  
$\eta = 1$. Comparison with Fig.~\ref{fig:f1} reveals that in both cases this time is about midway through the pump pulse. The detailed behavior of this switch-on
for different values of $N_0$ is illustrated in Fig.~\ref{fig:switch-on}. Note that in all cases, the switch-on of the output pulse is much more rapid than that of the pump
pulse illustrated in Fig.~\ref{fig:f1}.

The output pulse switch-on is also  much more rapid than is its subsequent decay, as seen in Fig.~\ref{fig:photon-no-1}. The function $x(t)$ still has a finite duration due to
the finite value of $N_0$, which implies that there will always be a last photon emitted, after which $x(t) = 0$. However, the switch-off is of less interest to us because the
high frequency behavior of the Fourier transform  of  $x(t)$   will depend primarily upon the rapid switch-on.

\subsection{An example of a specific laser transition}
\label{sec:transition}

So far, we have worked in units where the duration of the input pulse is set to one, $\tau_{\rm in} = 1$, and the output pulse is of the same order $\tau = \tau_{\rm out} \approx 1$.
However, it is useful to express these quantities in conventional units for specific cases. The parameter $C$ is the inverse of the radiative lifetime, $\tau_{\rm life}$, of the excited
state of the laser transition, or
\begin{equation}
 \tau_{\rm life} = \frac{\tau}{C}
 \end{equation} 
in general units. Let us consider the case of a Rhodamine 6G laser, for which the transition occurs at a wavelength of $\lambda \approx 570 \,{\rm nm}$
with a lifetime of $\tau_{\rm life} \approx 5.5 \times 10^{-9} \,{\rm s}$. (See Table~2.2 in Ref.~\cite{Svelto}.) The case $C=10^{-4}$ leads to $\tau \approx 5.5 \times 10^{-13}\, {\rm s}$,
and a pulse wave packet of length $\ell \approx 160\, \mu{\rm m} =  280 \,\lambda$.

\section{The Fourier transform of the envelope function.}
\label{sec:Fourier}

We can regard the output photon number flux, $r \, x(t)$, as the envelope function of the output pulse. Here we assume that the temporal duration of this pulse
is large compared to the characteristic period of the photons, so that the pulse may be viewed as a wavepacket containing a large number of oscillations under the
envelope function. In the specific example discussed above in Sec.~\ref{sec:transition}, the output pulse will contain about $280$ oscillations.
We are especially interested in the high frequency behavior of $\hat{x}(\omega)$,  the Fourier transform of $x(t)$. For the purposes of numerical
calculations, we compute the cosine and sin transforms separately, rather than the complex Fourier transform.
Let 
\begin{equation}
 F_C(\omega) = \int_0^\infty dt \, \cos(\omega t) \, x(t) \,,
 \end{equation}
and
 \begin{equation}
 F_S(\omega) = \int_0^\infty dt \, \sin(\omega t) \, x(t) \,.
 \end{equation}
We define
\begin{equation}
 F(\omega) = [F_C^2(\omega) + F_S^2(\omega)]^{1/2}
 \end{equation}
as our measure of the magnitude of the Fourier transform of the envelope function.

Our primary concern is whether $F(\omega)$ displays the asymptotic form in Eq.~\eqref{eq:FT-asy} , and if so, to estimate the associated value of the parameter $\alpha$.
This may be done by numerically computing $\log[F(\omega)]$, and then plotting this quantity as a function of $\omega$ on a log-log plot, which is
equivalent to a linear plot of $\log[\log(F)]$ as a function of $\log(\omega)$. An approximately straight line in  such a plot confirms the form of Eq.~\eqref{eq:FT-asy}, 
with  $\alpha$ being the magnitude of the slope.

\begin{figure}[htbp]
\includegraphics[scale=0.2]{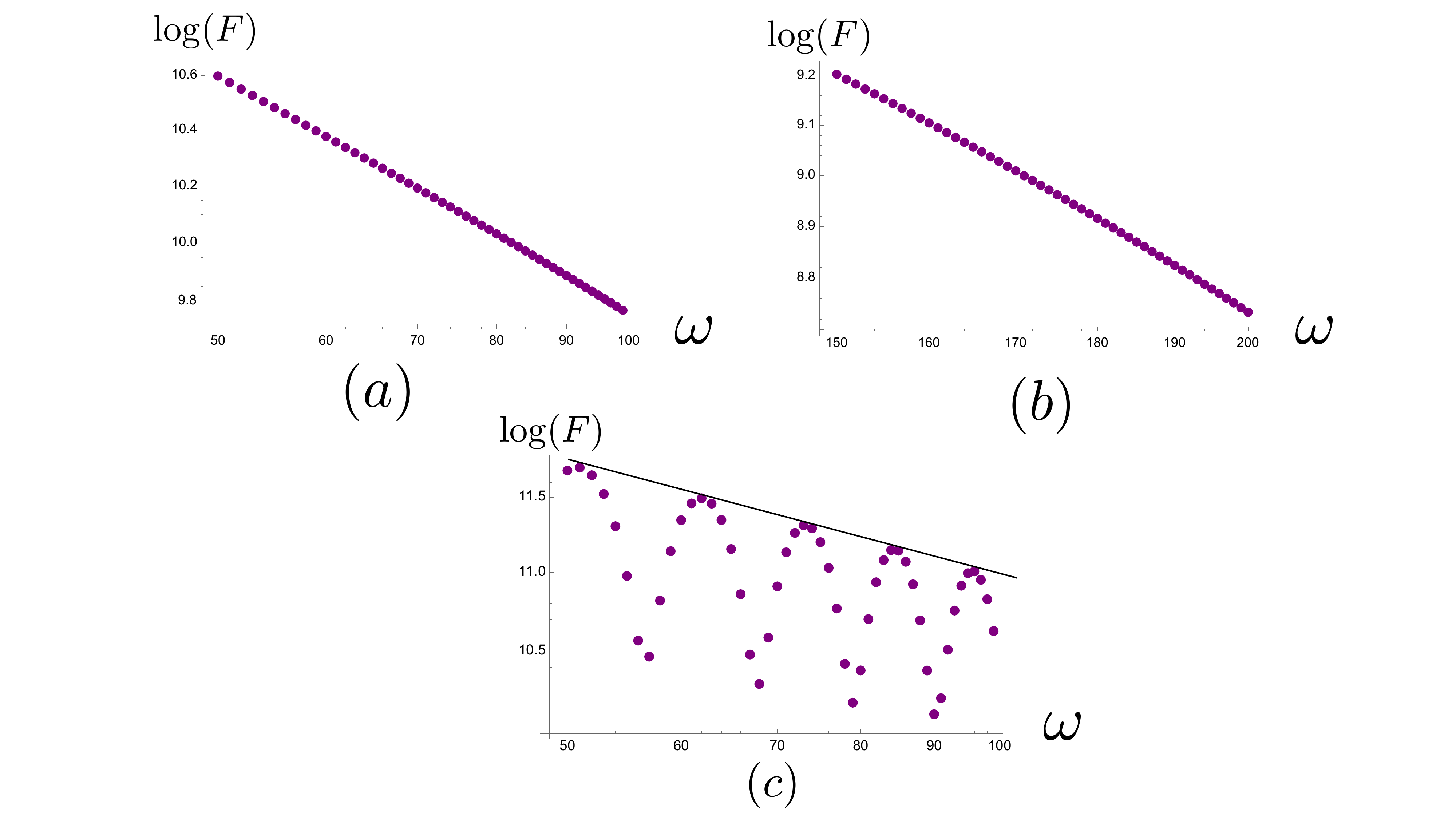}
\caption{The Fourier transforms of the envelope functions plotted in Fig.~\ref{fig:photon-no-1}, where $N_0 = 10^7$, $C=10^{-4}$, and $r = 2$, 
are plotted for the cases $\eta =1$ in parts (a) and (b) for different
frequency ranges, and for the case $\eta =0$ in part (c). All of these are log-log plots of $\log(F)$ as a function of $\omega$.  We find a straight line in  (a) and (b),
but a more complicated oscillatory behavior in (c). In the latter case, the value of $\alpha$ is estimated from the line connecting the maxima of the oscillations. Our
estimates for the slopes in each case are: $(a)\; \alpha \approx 0.12$, $(b)\; \alpha \approx 0.18$, $(c)\; \alpha \approx 0.097$.}
 \label{fig:alpha-fit1}
\end{figure}

This procedure is illustrated in Fig.~\ref{fig:alpha-fit1} for the two choices of $ x(t)$ depicted in Fig.~\ref{fig:photon-no-1}. Figure~\ref{fig:alpha-fit1}a, for the case
 with the absorption term, $\eta =1$, and $50 \leq \omega \leq 100$ , reveals a straight line with a slope of  $\alpha \approx (\log(10.6) - \log(9.8))/(\log(100) -\log(50)) \approx 0.11\,.$ 
 In a higher frequency range, $150 \leq \omega \leq 200$, shown in  Fig.~\ref{fig:alpha-fit1}b, we find $\alpha \approx 0.18$. 
 These results reveal two general patterns which we will find with other choices 
 of parameters: (1) The effective value of $\alpha$ of the output pulse tends to be much smaller than that of the input pulse, here $\alpha_{\rm in} = \frac{1}{2}$.
 (2) The fit value $\alpha$ can increase slowly as  $\omega$ increases.

The case without the absorption term, Fig.~\ref{fig:alpha-fit1}c, reveals oscillations in the Fourier transform, $\log[F(\omega)]$. In this case, we estimate the rate of decay
with increasing frequency from a straight line fit to the peaks of the oscillations, as illustrated. In this case, we find $\alpha \approx 0.097$, somewhat smaller
than in the case in  Fig.~\ref{fig:alpha-fit1}a, with the absorption term in the same frequency range.

\begin{table}[htbp]
\caption{Results for Varying $N_0$ with re-absorption ($\eta = 1$). 
Here $C = 10^{-4}$ and $r=2$.}
\centering
\begin{tabular}{|c c c c c|}
\hline\hline
 $N_0$ & \quad $\omega_{\rm min}$ & \quad $\omega_{\rm max}$ & \quad $\alpha \;{\rm estimate}$ &\quad $\omega_{\rm max}^\alpha$ \\
\hline
$10^7$ & 50 & 100 & 0.11 & 1.7 \\
             & 100 & 150 & 0.14 & 2.0 \\
             & 150 & 200 & 0.18 & 2.6 \\  
$10^8$ & 100 & 300 & 0.11 & 1.9 \\                       
              & 500 & 600 & 0.11 & 2.0 \\  
 $10^9$ & 50 & 500 & 0.13 & 2.2 \\                            
              & 1500 & 1600 & 0.19 & 4.1 \\  
  $10^{10}$ & 50 & 900 & 0.13 & 2.4 \\   
   $10^{11}$ & 50 & 900 & 0.12 & 2.3 \\                           
 $10^{12}$ & 50 & 900 & 0.11 & 2.1 \\  
\hline
\end{tabular}
\label{table:N0}
\end{table}

Some numerical results for $\alpha$ are given in Table~\ref{table:N0} as functions of $N_0$ in various frequency ranges, with $C = 10^{-4}$ and $r=2$.
Note that, as before, these results are all much smaller than was the value of  $\alpha$ for the pump pulse, $\alpha_{\rm in} = \frac{1}{2}$. The right-hand column gives
$\omega_{\rm max}^\alpha$ or each case, where $\omega_{\rm max}$ is the maximum frequency for the range in question. The significance of these numbers is that
$\exp(-\omega_{\rm max}^\alpha)$ is a measure of the probability distribution of a large fluctuation in the operator $u$ when averaged with the associated sampling function. In all
the cases  illustrated, this quantity exceed $1 \%$, which is vastly larger than the value of a Gaussian distribution for fluctuations of a similar magnitude compared to the variance. 
Recall that the values
of $\alpha$ in Table~\ref{table:N0} describe a probability distribution of the form of Eq.~\eqref{eq:Pu}, where $u$ lies in the range $\omega_{\rm min} \alt u \alt \omega_{\rm max}$
in $\tau = 1$ units. This follows from Eq.~\eqref{eq:omega-d}.

\begin{figure}[htbp]
\includegraphics[scale=0.2]{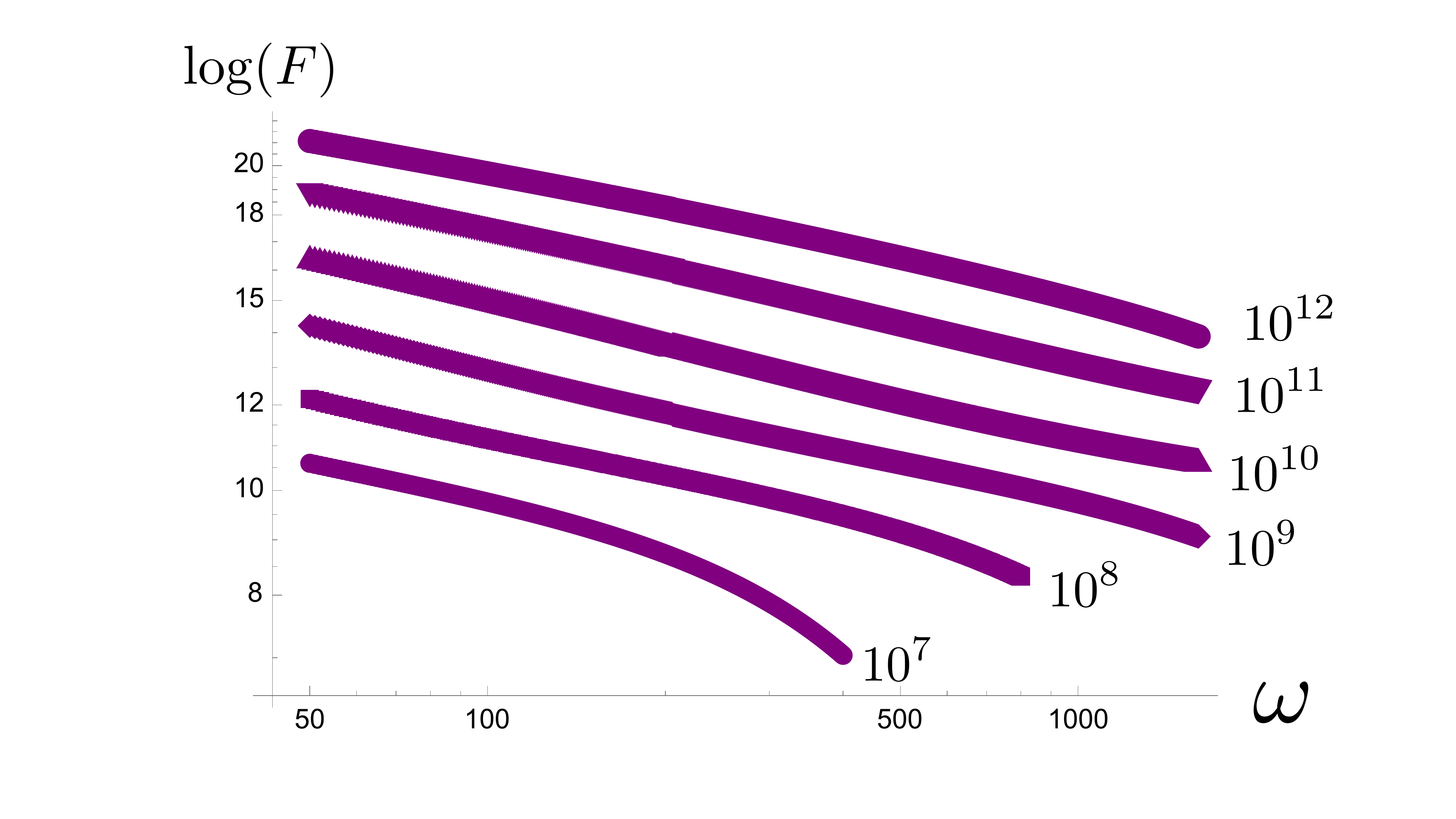}
\caption{Here $\log(F(\omega)$ is plotted for several values of $N_0$. Overall, these log-log plots approximate straight lines with their slopes being the
values of $\alpha$ reported in Table~\ref{table:N0}. The plots for the smaller values of $N_0$ tend to bend downward, leading to an effective value
of $\alpha$ which increases in the higher frequency ranges.}
 \label{fig:N0-plots}
\end{figure}

The corresponding log-log plots of  $\log(F)$ as a function of $\omega$ are given for various values of $N_0$ in Fig.~\ref{fig:N0-plots}. 

\begin{table}[htbp]
\caption{Results for Varying $N_0$ without re-absorption ($\eta = 0$), $C = 10^{-4}$ and $r=2$.}
\centering
\begin{tabular}{|c c c c c|}
\hline\hline
 $N_0$ & \quad $\omega_{\rm min}$ & \quad $\omega_{\rm max}$ & \quad Fractional oscillation amplitude   &   \quad $\alpha \;{\rm estimate}$  \\
\hline
$10^7$ & 51 & 96 & $17 \% $ & 0.097 \\
$10^8$ & 200 & 240 & $1.2 \%$  & 0.087 \\                       
 $10^9$ & 250 & 300 & $0.1 \%$ & 0.071 \\                            
 $10^{10}$ & 250 & 300 &  $< 0.1 \%$   & 0.061 \\   
   $10^{11}$ & 400 & 500 & $< 0.1 \%$ & 0.055 \\                           
 $10^{12}$ & 600 & 800 & $< 0.1 \%$ & 0.050 \\  
\hline
\end{tabular}
\label{table:N0-no absorb}
\end{table}

Table~\ref{table:N0-no absorb} gives some results for the case of no re-absorption ($\eta = 0$). The $N_0 = 10^7$ case was illustrated in Fig.~\ref{fig:alpha-fit1}(c), and shows large
oscillations. The amplitudes of these oscillations decrease rapidly as $N_0$ increases, and are not observed for $N_0 \geq 10^{10}$.  The results for $\alpha$ decrease with increasing
 $N_0$, and are somewhat smaller than the corresponding values with  re-absorption in Table~\ref{table:N0}.

\begin{table}[htbp]
\caption{Results for Varying $C$. Here $N_0 = 10^{7}$ and $r=2$. }
\centering
\begin{tabular}{|c c c  c|}
\hline\hline
 $C$ & \quad $\omega_{\rm min}$ & \quad $\omega_{\rm max}$   &   \quad $\alpha \;{\rm estimate}$  \\
\hline
$10^{-5}$ & 50 & 100      & 0.24 \\
                 &100 & 150    & 0.50 \\
$10^{-4.5}$ & 50 & 100   & 0.13 \\
                   &100 & 150    & 0.20 \\
  $10^{-4}$ & 50 & 100     & 0.11 \\
                   &100 & 150    & 0.12 \\   
  $10^{-3}$ & 50 & 100      & 0.16 \\
                   &100 & 150    & 0.15 \\   
 $10^{-2}$ & 50 & 100      & 0.22 \\
                   &100 & 150    & 0.23 \\
$10^{-1}$ & 50 & 100       & 0.23 \\
                   &100 & 150    & 0.26 \\                    
\hline
\end{tabular}
\label{table:C}
\end{table}

Table~\ref{table:C} gives some results on the variation of the parameter $C$ for the case $\eta = 1$, $N_0 = 10^{7}$ and $r=2$. We see that the values
of $\alpha$ tend to be smallest for $C \approx 10^{-4}$ and to increase for smaller or larger values. For $C \agt  10^{-4}$, the fit values of $\alpha$ are relatively 
independent of $\omega$ in the ranges displayed. However, for $C \alt  10^{-4}$, they increase rapidly with increasing $\omega$. This makes  $C \approx 10^{-4}$
the optimal choice for our purposes. We do not have simple explanation for this, but simply present it as an apparent numerical result.
 
\begin{table}[htbp]
\caption{Results for Varying $r$.  Here $N_0 = 10^{7}$ and $C = 10^{-4}$.}
\centering
\begin{tabular}{|c c c  c|}
\hline\hline
 $r$ & \quad $\omega_{\rm min}$ & \quad $\omega_{\rm max}$   &   \quad $\alpha \;{\rm estimate}$  \\
\hline
$4$ & 50 & 100      & 0.12 \\
$3$ & 50 & 100   & 0.13 \\
$2$ & 50 & 100     & 0.11 \\                        
$1.5$ & 50 & 100      & 0.12\\
 $1$ & 50 & 100      & 0.11 \\
$0.8$ & 50 & 100       & 0.12 \\
$0.4$ & 50 & 100       &  0.12 \& 0.11    \\
$0.2$ & 50 & 100       &  0.13 \& 0.066    \\                
\hline
\end{tabular}
\label{table:r}
\end{table}

Table~\ref{table:r} gives some results on the variation of the parameter $r$ for the case $\eta = 1$, $N_0 = 10^{7}$ and  $C = 10^{-4}$.
Recall that this parameter describes the rate at which photons leave the cavity, with larger $r$ corresponding to a greater rate. We see
that  $\alpha$ is nearly independent  of $r$, except when $r \alt 0.4$. In this case we find two populations of photons with different values
of $\alpha$. This may be due to the fact that smaller values of $r$ correspond to a greater reflectivity of the partially reflecting mirror through which
the photons exit the cavity. This enhances the probability of multiple reflections, which may be linked to the presence of different photon populations.

\section{Possible Applications}
\label{sec:apps}

\subsection{Density Fluctuations in Condensed Matter}
\label{sec:density}

A liquid or solid can undergo mass density fluctuations due to zero point motion of the atoms in the phonon vacuum state. This effect was
discussed in Refs.~\cite{FS09,WF20}. These references explicitly considered the case of a liquid, but the analysis  applies to amorphous
solids and also to crystalline solids if the relevant phonon wavelengths are large compared to the lattice spacing. The typical magnitude of
the density fluctuations and their observability in light scattering experiments was discussed. The second quantized phonon field operator
is formally equivalent to that for the time derivative of a massless scalar field in relativistic field theory, but with the speed of light replaced
by the speed of sound~\cite{Unruh,FF04}. 

Here a typical magnitude fluctuation is one described by the variance of the phonon operator.  However, if the light scattering involves an isolated
laser pulse of the type described in the present paper, then the probability of a much larger fluctuation can be significant. Let the probability
distribution for a large fluctuation be given by Eq.~\eqref{eq:P-greater}, where  $u$ denotes the ratio of a given density fluctuation to its variance.
We may extend some results in Ref.~\cite{WF20} for light scattering in liquid ${\rm He}^3$, and  rewrite Eq.~(24) in that reference as
\begin{equation}
\frac{\Delta n_s}{n_T} \approx 4.6 \, u \,
\left(\frac{160\mu\text{m}}{\ell}\right)^5\left(\frac{200 \text{m/s}}{c_s}\right)^{3/2}\left(\frac{\lambda_0}{570 \text{nm}}\right)^4 \left(\frac{1 K}{T}\right)\,.
\label{eq:ns-nT}
\end{equation}
Here $\Delta n_s$ is the expected number of photons scattered by a localized zero point density fluctuation, as illustrated in Fig.~2 of Ref.~\cite{WF20}, from a pulse of length $\ell$
and mean wavelength $\lambda_0$. The mean number of photons scattered from this pulse by thermal fluctuation at temperature $T$ is $n_T$.  

Note that Ref.~\cite{WF20}
focused on the case of typical fluctuations, $u \approx O(1)$.  Light scattering experiments in various materials~\cite{Stephen69,VP76,Eramo,FS11} can be viewed as evidence for zero point 
motion, but to our knowledge, experiments using localized pulses have not yet been performed.
 In any case, the observation of large density fluctuations, those with $u \gg 1$, would be of great interest
We can see that in this case, zero point fluctuations can easily dominate thermal effects. Equation~\eqref{eq:ns-nT} applies specifically to  liquid ${\rm He}^3$, with an
index of refraction of $1.026$, but provides a reasonable order of magnitude estimate to other materials and higher temperatures. Based upon our results in 
Table.~\ref{table:N0}, for example, it appears that density fluctuations with $u$ of the order of $10^2$ or larger may be observable. In this case, zero point  fluctuations
could become comparable to thermal effects even at room temperature. This provides a motivation for the further study of the Fourier transforms of laser pulse envelopes.
However, a careful estimate of the probability of a large density fluctuation with a given value of $u$ will require a calculation of the values of the constants
$c_0$ and $b$ in Eqs.~\eqref{eq:Pu} and \eqref{eq:P-greater}, which is also a topic for future work. 

\subsection{Radiation Pressure Fluctuations on Electrons}
\label{sec:RadPress}

A more speculative possibility is the creation of isolated pulses of gamma rays by backscattering of optical photons in a laser pulse from high energy 
electrons~\cite{Milburn63,Milburn65}. Such a pulse of gamma rays might produce observable vacuum radiation pressure effects on other electrons~\cite{F24}.
However, more work is needed to explore this possibility. In particular, it is not clear whether the value of the parameter $\alpha$ for the gamma ray pulse can be
as small as that for the optical laser pulse.

\section{Summary}
\label{sec:final}

In this paper, we have presented results on the Fourier transform of the time envelope function for a laser pulse of finite duration. 
Section~\ref{sec:intro} gave some motivation for this investigation. Here we reviewed results of the probability distributions for the 
zero point fluctuations of quadratic operators which have been averaged  in time by such a pulse. The key result is that  these are 
non-Gaussian distributions which decay as an exponential of a fractional power, $\alpha < 1$. Furthermore, an exponential of the same 
power also appears in the Fourier transform of the envelope function. 
[Compare  Eq.~\eqref{eq:FT-asy} with Eqs.~\eqref{eq:Pu} and \eqref{eq:P-greater}.] 
In Sec.~\ref{sec:rate} we presented a model, based upon rate equations, for the generation of finite
duration pulses, and analyzed the  Fourier transform of the output envelope function in Sec.~\ref{sec:Fourier}. Our results for $\alpha$  are 
remarkably small, typically in the range $0.1 \alt \alpha \alt 0.2$. These results are significantly smaller than the value assumed for the 
input (pump) pulse of $\alpha_{\rm in} = \frac{1}{2}$. The latter
is a value which can be obtained in the switch-on of simple electrical circuits~\cite{FF15}. 
The slow decay of the  Fourier transform at high frequencies is likely linked
to the very short switch-on times of the output pulse illustrated in Fig.~\ref{fig:switch-on}.

The relatively small values of  $\alpha$ obtained imply a significant probability for zero point fluctuations of quadratic operators large compared to the variance. Two physical examples 
of such operators were treated in Sec.~\ref{sec:apps}. Of particular interest are large density fluctuations in condensed matter, which may be observable in light scattering 
experiments using pulses of the type treated here.

\begin{acknowledgments} 
LHF would like to thank Chris Fewster  for useful conversations, especially on the material in Appendix~\ref{sec:App A}.
This work was supported in part  by the National Science Foundation under Grant PHY-2207903.
\end{acknowledgments}

\appendix
 \section{Bounds on the decay rate of Fourier transforms}
 \label{sec:App A}

\begin{figure}[htbp]
\includegraphics[scale=0.2]{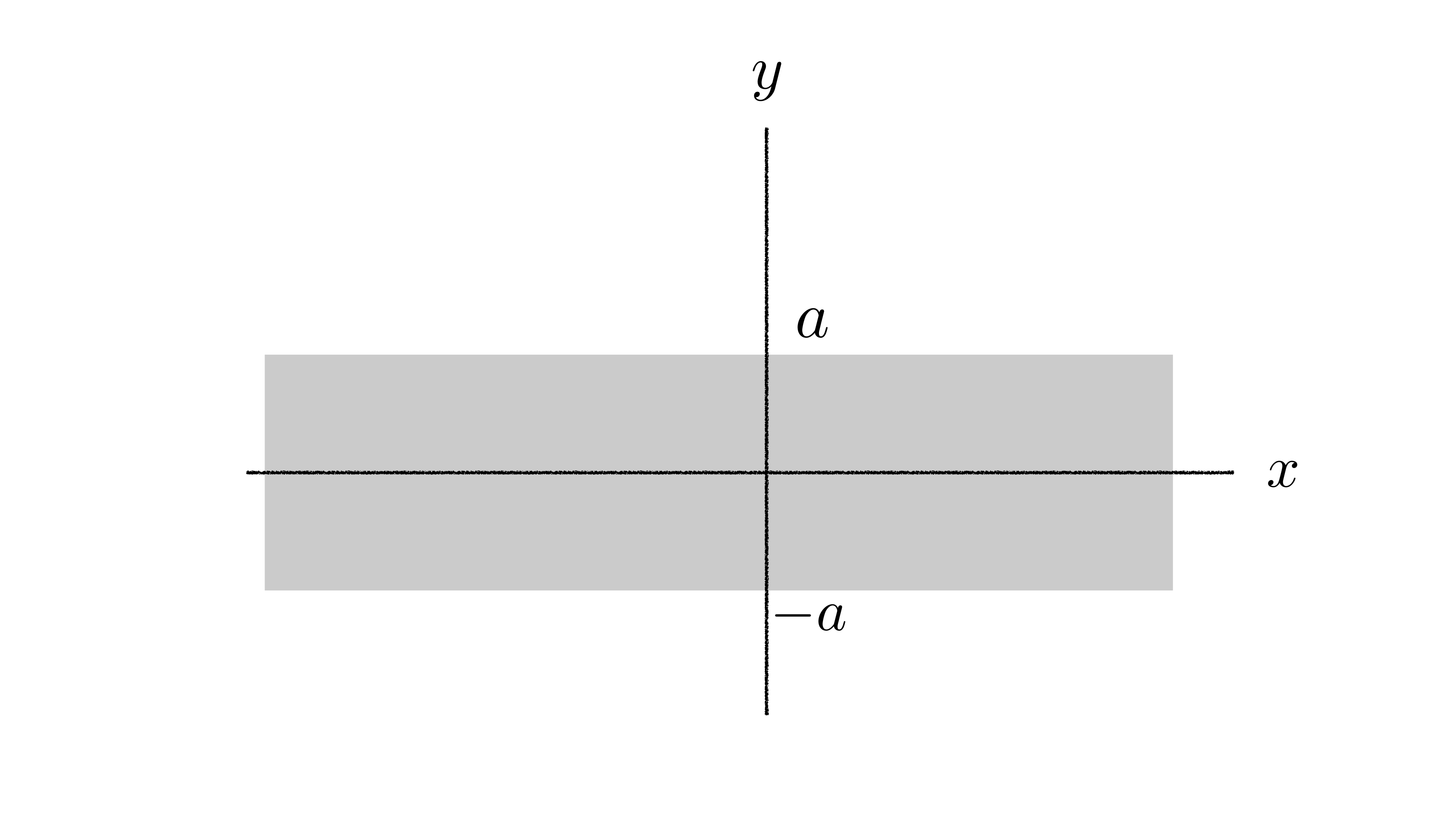}
\caption{The region of analyticity of the function $f(t)$ in the complex plane is illustrated, under the assumption of Eq.~\eqref{eq:decay-postulate}. 
Here $t = x +i y$, and $f(t)$ is analytic in the strip $|y| < a$.}
\label{fig:analytic-strip}
\end{figure}

Here we wish to prove the theorem that an infinitely differentiable function with compact support must have a Fourier transform which decays more
slowly than an exponential function but faster than any power.  Let $f(t)$ be such a function whose Fourier transform is $\hat{f}(\omega)$, so that
\begin{equation}
 f(t) = \frac{1}{ 2 \pi} \, \int_{-\infty}^{\infty} d\omega \, \hat{f}(\omega) \, {\rm e}^{i \omega t}\,.
 \label{eq:Fourier-rep}
 \end{equation}
 If we take an arbitrary number of derivatives of Eq.~\eqref{eq:Fourier-rep} with respect to $t$, the integral must still converge if $f(t)$ is
 infinitely differentiable. This requires that $\hat{f}(\omega)$ fall faster than any power as $|\omega| \rightarrow \infty$. Next we address 
the lower bound on this rate of decay. Let $t = x +i y$. Then
\begin{equation}
 f(x +i y) = \frac{1}{ 2 \pi} \, \int_{-\infty}^{\infty} d\omega \, \hat{f}(\omega) \, {\rm e}^{i \omega x}\,  {\rm e}^{- y}  .
 \end{equation}
 First assume that $\hat{f}(\omega)$ does decay at least exponentially for large $|\omega|$:
  \begin{equation}
 |\hat{f}(\omega)| \leq C\,  {\rm e}^{- a |\omega| }
 \label{eq:decay-postulate}
 \end{equation}
for some constants $a > 0$ and $C$. This implies that
\begin{equation}
 |f(x +i y)|   \leq C\,  \, \int_{-\infty}^{\infty} d\omega \,  {\rm e}^{ -a\, |\omega| }\,  {\rm e}^{-\omega y}\, .
 \end{equation}
This integral is absolutely convergent in the strip defined by $|y| < a$, illustrated in Fig.~\ref{fig:analytic-strip}. Hence the function  $f(t)$
is analytic in this strip. However, analyticity is inconsistent with the assumption of compact support because the only analytic function which vanishes in
a finite region is the trivial case which is identically zero everywhere.  Consequently,  Eq.~\eqref{eq:decay-postulate} cannot be correct.
We conclude that $|\hat{f}(\omega)|$ must decay more slowly than any exponential for large $\omega$.

\end{document}